\documentclass[12pt]{article}
\usepackage{amssymb}
\usepackage{a4}
\usepackage{amsmath}
\usepackage{epsfig}
\usepackage
{hyperref}
\usepackage{color}
\definecolor{rltred}{rgb}{1,0.0,.0}
\definecolor{rltgreen}{rgb}{0.5,0.5,0.5}
\definecolor{rltblue}{rgb}{0,0,0.75}
\newcommand{\IE}{\mathbb{E}}
\newcommand{\IP}{\mathbb{P}}
\newcommand{\R}{\mathbb{R}}
\newcommand{\C}{\mathbb{C}}

\newcommand{\Z}{\mathbb{Z}}
\newcommand{\ord}{\mathcal{O}}
\newcommand{\Ai}{\text{Ai}}

\begin{document}
\bibliographystyle{unsrt} 
\title{Exact scaling functions for
  one-dimensional stationary KPZ growth} 
\author{Michael Pr\"ahofer
  and Herbert Spohn,\\
\small  Zentrum Mathematik and Physik Department, TU M\"unchen,\\
\small D-80290 M\"unchen, Germany\\
\small emails: {\tt praehofer@ma.tum.de}, {\tt
  spohn@ma.tum.de}\medskip\\
\small 
With deep appreciation dedicated to Giovanni Jona-Lasinio at\\[-1mm]
\small
 the
occasion of his 70th birthday.}
\date{} 

\maketitle
\begin{abstract}
  We determine the stationary two-point correlation function of the
 one-dimensional KPZ equation through the scaling limit of a solvable
 microscopic 
 model, the polynuclear growth model. The equivalence to a directed
 polymer problem with specific boundary conditions allows one to express
 the corresponding scaling function in terms of the solution to a
 Riemann-Hilbert  problem related to the Painlev\'e II equation. We
 solve these equations numerically with very high precision and
 compare our, up to numerical rounding exact, result with the
 prediction of Colaiori and Moore 
 \cite{moore01:_numer_solut_mode_coupl_equat} obtained from the mode
 coupling approximation. 
\end{abstract}

\section{Introduction}
\setcounter{equation}{0}
\label{sec:intro}

In their well-known work \cite{kardar86:_dynam} Kardar, Parisi, and
Zhang argue that surface growth through random ballistic deposition
can be modeled by a stochastic continuum equation, which in the case
of a one-dimensional substrate reads
\begin{equation}
  \label{eq:KPZ1d}
  \partial_th=\textstyle\frac12\lambda(\partial_xh)^2
  +\nu\partial_x^2h+\eta. 
\end{equation}
Here $h(x,t)$ is the height at time $t$ at location $x$ relative to a
suitable reference line. $\eta(x,t)$ is space-time white noise of
strength $D$,
$\langle\eta(x,t)\eta(x',t')\rangle=D\delta(x-x')\delta(t-t')$, and
models the randomness in deposition. $\nu\partial_x^2h$ is a not
further detailed smoothening mechanism. The important insight of
\cite{kardar86:_dynam} is to observe that the growth velocity is
nonlinear, in general, and is relevant for the large scale properties
of the solution to (\ref{eq:KPZ1d}). To simplify, the growth velocity
is expanded in the slope. The first two terms can be absorbed through
a suitable choice of coordinate frame. The quadratic nonlinearity in
(\ref{eq:KPZ1d}) is relevant and higher orders can be ignored, unless
$\lambda=0$.

The one-dimensional KPZ equation (\ref{eq:KPZ1d}) is regarded as
exactly solved in the usual terminology. In fact, what can be obtained
is the dynamic scaling exponent $z=3/2$ 
\cite{forster77:_large,kardar86:_dynam,beijeren85:_exces}. No other
universal quantity has been computed exactly so far. In our
contribution we will improve the situation and explain how to extract
the scaling function for the stationary two-point function. A few other
universal quantities can be computed as well. But they have been
discussed already elsewhere
\cite{praehofer01:_curren,praehofer00:_univer,praehofer01:_scale_invar_png_dropl_airy_proces}.

In \cite{beijeren85:_exces} a mode-coupling equation for the two-point
function is written down, in essence following the scheme from
critical dynamics and kinetic theory. At the time only $z=3/2$ and a
few qualitative properties could be extracted from the mode-coupling
equation. In \cite{frey96:_mode_burger} this equation is solved
numerically. Such computations are repeated in
\cite{moore01:_numer_solut_mode_coupl_equat} with greatly improved
precision and using a more convenient set of coordinates. Thus for the
1D KPZ equation we are in the unique position of an exact solution and
an accurate numerical solution to the mode-coupling equation with {\it
  no} adjustable parameters. As will be explained below, given the
uncontrolled approximation, mode-coupling does surprisingly well.

To attack (\ref{eq:KPZ1d}) directly does not seem to be feasible, a
situation which is rather similar to the one for two-dimensional
models in equilibrium statistical mechanics. For example, the
Ginzburg-Landau 
$\phi^4$-theory is given through the (formal) functional measure
\begin{equation}
\label{eq:phifour}
  Z^{-1}\prod_{x\in\R^2}d\phi(x)
  \exp\Big[-\int d^2x\Big((\nabla\phi)^2+g\phi^2+\phi^4\Big)\Big]
\end{equation}
for the scalar field $\phi$. (\ref{eq:phifour}) is not the proper
starting point for computing the exact two-point scaling function at
the critical coupling $g_c$. Rather one discretizes through the
lattice $\Z^2$ and replaces the $\phi$-field by Ising spins
$\pm1$. Then, following e.g.~\cite{wu76:_ising}, the scaling function at and
close to criticality can be obtained. By universality this scaling
function is the one of (\ref{eq:phifour}). (While
certainly true, to establish universality is difficult and carried out
in a few cases only \cite{spencer00}.) In the same spirit we replace
(\ref{eq:KPZ1d}) by a discrete model, where the most convenient choice
seems to be the polynuclear growth (PNG) model.

Before explaining the PNG model let us review the standard scaling
theory for (\ref{eq:KPZ1d}). If the initial conditions $h(x,0)$ of the
KPZ equation are
distributed according to two-sided Brownian motion, then formally the
distribution of $h(x,t)-h(0,t)$ is again two-sided Brownian
motion. Therefore it is natural to define the stationary time correlation 
\begin{equation}
  \label{eq:statcorr}
  C(x,t)=\big\langle\big(h(x,t)-h(0,0)
  -t\langle\partial_th\rangle\big)^2\big\rangle,
\end{equation}
where from the height difference the average displacement
is subtracted. By assumption
\begin{equation}
  \label{eq:equaltime}
  C(x,0)=A|x|
\end{equation}
with roughness amplitude $A=D/\nu$ to ensure stationarity in time. If
$z=3/2$, then $C(x,t)$ scales as 
\begin{equation}
  \label{eq:generalscalingform}
  C(x,t)\propto t^{2/3}g(const\cdot x/t^{2/3}),\quad \mbox{as
  $x,t\to\infty$} 
\end{equation}
with a universal scaling function $g(y)$ having the asymptotics
$g(y)\to c_0>0$ for $y\to0$ and $g(y)\sim c_\infty|y|$ for
$|y|\to\infty$.  In order to define $g$ as a dimensionless function we
fix the proportionality constants in (\ref{eq:generalscalingform}) as
appropriate combinations of $\lambda$ and $A$,
\begin{equation}
  \label{eq:dimensionlessscalingform}
  g(y)=\lim_{t\to\infty}\frac{C\big((2\lambda^2A\,t^{2})^{1/3}y,t\big)}
  {(\frac12\lambda A^2t)^{2/3}},
\end{equation}
where the particular choice of numerical prefactors is in principle
arbitrary. The factor $2^{-2/3}$ in the denominator
is chosen in order to conform with the convention for the GUE
Tracy-Widom distribution \cite{tracy94:_level_airy_kernel}. The factor
$2^{1/3}$ in the argument of the numerator differs from the convention
used by Baik and Rains \cite{baik00:_limit} by a factor $2$ but
conforms with the definition of the closely related Airy process
\cite{praehofer01:_scale_invar_png_dropl_airy_proces} and has the
further advantage to absorb a lot of prefactors in the equations
defining $g(y)$. Note however, that the exponents for the parameters 
$\lambda\big[\frac{x^2}{th}\big]$, $\nu\big[\frac{x^2}{t}\big]$, and
$D\big[\frac{h^2x}{t}\big]$ are fixed uniquely by dimensional
reasoning.

We remark that the slope $\partial_xh(x,t)$ is space-time stationary
in the usual sense. For fixed $t$, $x\mapsto\partial_xh(x,t)$ is white
noise with strength $A$. Since $\langle\partial_xh\rangle=0$, the
standard 2-point function is 
\begin{equation}
  \label{eq:derivtwopt}
  \langle\partial_xh(0,0)\partial_xh(x,t)\rangle=
  \textstyle\frac12\partial_x^2C(x,t).
\end{equation}
This relation and the asymptotic behavior of $g$, $g(y)/|y|\to2$ as
$y\to\infty$,
motivates the definition of a second scaling function,
\begin{equation}
  \label{eq:fdef}
  f(y)=\textstyle\frac14g''(y),
\end{equation}
which by definition has integral normalized to one and which will be
shown to be positive in the next section.

In the sequel we will analyze the distribution function for the height
differences in the stationary PNG model. As shown in
\cite{baik00:_limit}, they can be represented in terms of certain
orthogonal polynomials, which lead to recursion relations connected to
the Painlev\'e II differential equation
\cite{baik01:_rieman,periwal90:_unitar}. The asymptotic analysis is
carried out in \cite{baik00:_limit}. Our own contribution is twofold:
(i) We observe that the stationary PNG model maps to a last passage
percolation with boundaries \cite{praehofer00:_univer}. (ii) The
expressions in \cite{baik00:_limit} are given in terms of certain
differential equations and the extraction of the scaling function $g$
requires a careful numerical integration. This is one central point of
our article. We will provide then plots of the structure function and
give a comparison with the mode-coupling theory.

\section{The polynuclear growth model}
\setcounter{equation}{0}
\label{sec:PNG}

The polynuclear growth (PNG) model is a model for layer-by-layer
growth through deposition from the ambient atmosphere. The surface is
parameterized by a time dependent integer-valued height function
$h(x,t)$, $t\in\R$, above a one-dimensional 
substrate, $x\in\R$. Thus the height function consists of terraces
bordered by steps of unit height. The up-steps move to the left and
the down-steps to the right with speed $1$. Steps disappear upon
collision. In addition to this deterministic dynamical rule new
islands of unit height are nucleated randomly with
space-time density $2$ on top of already
existing terraces. The corresponding stochastic process $h(x,t)$
is well defined even in infinite volume (cf. \cite{seppaelaeinen96}
for the closely related Hammersley particle process).

Of interest to us here is the stationary growth process, which means
that the slope $\partial_xh(x,t)=\rho(x,t)$ is stationary in
space-time. One can think of $\rho(x,t)$ as the density of a
particle/antiparticle 
process. The particles are located at the up-steps and thus move with
velocity $-1$, the antiparticles are located at the down-steps 
and move with velocity $1$. Upon collision particle/antiparticle
pairs annihilate. In addition, with space-time density $2$, a
particle/antiparticle pair is created with the particle moving to the
left, the antiparticle to the right. To make $\rho$ stationary, one
prescribes at $t=0$ up-steps 
Poisson distributed with density $\rho_+$ and down-steps independently
Poisson distributed with intensity $\rho_-$ such that
\begin{equation}
  \label{eq:rhocond}
  \rho_+\rho_-=1.
\end{equation}
This measure for steps is stationary under the PNG dynamics. The mean
slope is given by
\begin{equation}
  \label{eq:meanslope}
  u=\rho_+-\rho_-=\langle\partial_xh(x,t)\rangle,
\end{equation}
which is the only remaining free parameter. For fixed $t$, $x\mapsto
h(x,t)-h(0,t)$ is a (two-sided) random walk with rate $\rho_\pm$ for a
jump from $n$ to $n\pm1$. It has average $u$ and variance
$\rho_++\rho_-$, which implies for the roughness amplitude
\begin{equation}
  \label{eq:roughampl}
  A(u)=\sqrt{4+u^2}.
\end{equation}
For the growth velocity one obtains
\begin{equation}
  \label{eq:growthvelo}
  v(u)=\langle\partial_th\rangle=\rho_++\rho_-=\sqrt{4+u^2}.
\end{equation}

Given $\rho(x,t)$ the height $h(x,t)$ is determined only up to a
constant which we fix as $h(0,0)=0$. To emphasize that only height
differences count, $h(0,0)$ is sometimes kept in the formulas.

The stationary process with slope $u$ transforms to the stationary
process with slope $0$ through the Lorentz transformation
\begin{equation}
  \label{eq:lorentztransf}
  x'=(1-c^2)^{-1/2}(x-ct),\quad
  t'=(1-c^2)^{-1/2}(t-cx),
\end{equation}
with the speed of ``light'' equal to $1$ and the velocity parameter
$c=-u/\sqrt{4+u^2}$. 
Thus it suffices to restrict ourselves to $u=0$ which we do from now
on. In particular $\rho_+=1=\rho_-$.
$\langle\,\cdot\,\rangle$ and $\IE$ refer to the stationary 
density field at slope $u=0$.

The central objects are the height-height correlation
\begin{equation}
  \label{eq:pngcorr}
  C(x,t)=\langle(h(x,t)-h(0,0)-2t)^2\rangle
\end{equation}
and the closely related two-point function for the density,
\begin{equation}
  \label{eq:pngtwopt}
  S(x,t)=\langle\rho(x,t)\rho(0,0)\rangle.
\end{equation}
They are related as
\begin{eqnarray}
  \label{eq:twopointdensity}
  \textstyle\frac12\partial_x^2C(x,t)&=&
  \textstyle\frac12\partial_x^2\IE\Big(\big(h(x,t)-h(0,0)-2t\big)^2\Big)
  \nonumber\\
  &=&\partial_x\IE\Big(\rho(x,t)\big(h(x,t)-h(0,0)-2t\big)\Big)
  \nonumber\\
  &=&\partial_x\IE\Big(\rho(0,t)\big(h(0,t)-h(-x,0)-2t\big)\Big)
  \nonumber\\
  &=&\IE\big(\rho(0,t)\rho(-x,0)\big)
  \nonumber\\
  &=&S(x,t).
\end{eqnarray}

\begin{figure}[tbp]
  \begin{center}
    \includegraphics[width=10cm]{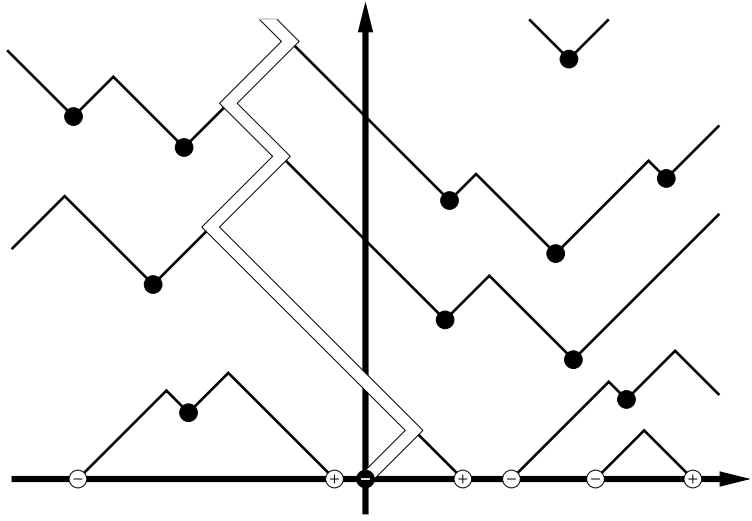}
    \caption{\it The trajectory of a second-class particle.}
    \label{fig:secondclass}
  \end{center}
\end{figure}

The height correlation is convex, equivalently
\begin{equation}
  \label{eq:convexity}
  S(x,t)\geq0.
\end{equation}
To prove this property we show that the structure function $S(x,t)$
can be regarded as the transition probability for a second class
particle starting at the origin. Its initial velocity is $\pm1$ with
probability $\frac12$, as for the ``first-class'' 
up/down-steps. In contrast to an ordinary step the second class
particle is never destroyed
upon colliding with another step. Rather it eats up the
step encountered and, by reversing its own direction of motion,
continues along the trajectory of the absorbed step, cf.~Figure
\ref{fig:secondclass}. 
Let $\rho(x,t)$ be a given realization of the PNG process. The second
class particle is added as
\begin{equation}
  \label{eq:secondclass}
  \rho^{(\sigma)}(x,0)=\rho(x,0)+\sigma\delta(x),\quad \sigma=\pm1.
\end{equation}
$\rho^{(\sigma)}(x,0)$ evolves to $\rho^{(\sigma)}(x,t)$ with
nucleation events identical to the one for $\rho(x,t)$.
By construction, if $X_t$ denotes the position of the second class
particle at time $t$,
\begin{equation}
  \label{eq:difference}
  \rho^{(\sigma)}(x,t)-\rho(x,t)=\sigma\delta(x-X_t).
\end{equation}
Noting that by the Poisson property $\rho^{(\sigma)}(x,0)$ is given by
$\rho(x,0)$ conditioned on the presence of either an up-step
($\sigma=+1$) or down-step ($\sigma=-1$) at the origin, we obtain
\begin{eqnarray}
  \label{eq:2ndclass}
  0\leq p_t(x)&=&\frac12\sum_{\sigma=\pm1}
  \IE\Big(\sigma\big(\rho^{(\sigma)}(x,t)-\rho(x,t)\big)\Big)=
  \frac12\sum_{\sigma=\pm1}\IE\big(\sigma\rho^{(\sigma)}(x,t)\big)
    \nonumber\\
  &=&\lim_{\delta\searrow0}\frac12\sum_{\sigma=\pm1}
  \sigma\IE\Big(\rho(x,t)\Big|\int_{-\delta}^\delta\rho(y,0)\,dy=\sigma\Big)
  \nonumber\\
  &=&\lim_{\delta\searrow0}\frac12\sum_{\sigma,\sigma'=\pm1}
  {\sigma\sigma'}\,\,
  \frac{\IP\big\{\int_{-\delta}^\delta\rho(x+y,t)\,dy=\sigma',
   \int_{-\delta}^\delta\rho(y,0)\,dy=\sigma\big\}}
  {{2\,\delta}\,\IP\big\{\int_{-\delta}^\delta\rho(y,0)\,dy=\sigma\big\}}
  \nonumber\\[2mm]
  &=&\textstyle\frac12\IE\big(\rho(x,t)\rho(0,0)\big)=\textstyle\frac12S(x,t).
\end{eqnarray}

For arbitrary slope $u$ the normalization of $S(x,t)$ would be given by
$v(u)=\sqrt{4+u^2}$ and the mean of $p_t(x)$ evolves along the
characteristics of the 
macroscopic evolution equation $\partial_tu=-\partial_xv(u)$. Thus
\begin{equation}
  \label{eq:meanbehavior}
  \int S(x,t)dx=\sqrt{4+u^2},\quad\mbox{and}\quad\int x\,S(x,t)dx=-t\,u.
\end{equation}

\section{The distribution functions for the height differences}
\label{sec:heightdistribution}
For the PNG model the distribution function for the height difference
$h(x,t)-h(0,0)$ satisfies certain recursion relations, which are the
tool for analyzing the scaling limit when $t\to\infty$ and
$x=y\,t^{2/3}$ with $y=\ord(1)$. The second moments yield $C(x,t)$ and
therefore by (\ref{eq:twopointdensity}) also $S(x,t)$.

Since the nucleation events are Poisson, $h(x,t)-h(0,0)$ depends only
on the events in the backward light cone $\{(x',t')\in\R^2,\,0\leq
t'\leq t,\,|x-x'|\leq t-t'\}$ and the initial conditions at $t=0$. Along
the line $\{x'=t'\}$ the down-steps are Poisson distributed with line
density $\sqrt{2}$ and correspondingly for the up-steps along the line
$\{x'=-t'\}$. This property can be deduced from the uniqueness
  of the stationary state at given slope and the Lorentz invariance
  (\ref{eq:lorentztransf}) in the limit $c\to\pm1$. Thus
$h(x,t)-h(0,0)$ is determined by the nucleation 
events in the rectangle $R_{x,t}=\{(x',t')\in\R^2,\,
|x'|\leq|t'|,\,|x-x'|\leq t-t'\}$ together with the said boundary
conditions. $h(x,t)-h(0,0)$ can be reexpressed as a directed last passage
percolation according to the following rules: Inside $R_{x,t}$ there
are Poisson points with density $2$. Along the two lower edges of
$R_{x,t}$ there are independently Poisson points with line density
$\sqrt{2}$. A directed passage from $(0,0)$ to $(x,t)$ is given
through a directed 
path (polymer). It is a piecewise linear path in the plane, starts at
$(0,0)$ and ends at $(x,t)$, alters its direction only at Poisson
points, and is time-like in the sense that for any two points
$(x',t')$, $(x'',t'')$ on the path one has $|x'-x''|\leq|t'-t''|$.
Note that, once the directed path leaves one of the lower edges to
move into the bulk, it can never return. By definition the
length of a directed path equals the number of Poisson points
traversed. With these conventions
\begin{equation}
\label{incrsubsequ}
  h(x,t)-h(0,0)=\,\mbox{maximal length of a directed path from $(0,0)$
  to $(x,t)$}.
\end{equation}
We remark that in general there are several maximizing paths, their
number presumably growing exponentially with $t$.

Under the Lorentz transformation (\ref{eq:lorentztransf}) the
distribution for the height differences (\ref{incrsubsequ}) does not
change. Therefore, we might as well transform $R_{(x,t)}$ to a
square. By an additional overall scaling by $\sqrt{2}$ one arrives at
a $v\times v$ square, $v=\sqrt{t^2-x^2}$, with bulk density $1$ and the 
line densities $\alpha_-=\alpha=\sqrt{(t-x)(t+x)}$ for the lower left,
resp. $\alpha_+=1/\alpha$ for the lower right edge. 
In this way we have recovered precisely the setting in
\cite{baik00:_limit,baik01:_rieman} with $t$ replaced by $v$. Baik and
Rains derive an explicit expression for the height distribution in
terms of Toeplitz determinants, which can be further simplified by
means of corresponding orthogonal polynomials. 

Let us state the result
for the distribution function of $h(x,t)-h(0,0)$,
\begin{eqnarray}
  \label{eq:differencedistribution}
  F_{x,t}(n)&=&\IP\{h(x,t)-h(0,0)\leq
  n\}
  \nonumber\\
  &=&G_n(\alpha)F(n)-G_{n-1}(\alpha)F(n-1).
\end{eqnarray}
For fixed $v$ the functions $g$ and $F$ are given in terms of the
monic polynomials $\pi_n(z)=z^n+\ord(z^{n-1})$, which are pairwise
orthogonal on the unit circle $|z|=1$ with respect to the weight
$e^{v(z+z^{-1})}$. Their norm $N_n$ is given by 
\begin{equation}
  \label{eq:ortho}
  \langle\pi_n,\pi_m\rangle=\delta_{n,m}N_n
\end{equation}
with $\langle
p,q\rangle=\oint p(z)q(z^{-1})e^{v(z+z^{-1})}(2\pi i z)^{-1}{dz}$.
One has 
\begin{equation}
  \label{eq:Fndef}
  F(n)=e^{-v^2}\prod_{k=0}^{n-1}N_k,
\end{equation}
where $F(n)$ itself is the distribution function of the maximal length
of a directed path in the case $\alpha_+=0=\alpha_-$. Thus
$\lim_{n\to\infty}F(n)=1$ and
\begin{eqnarray}
  \label{eq:ydef}
  G_n(\alpha)&=&e^{-v(\alpha+\alpha^{-1})}N_n\sum_{k=0}^n
  N_k^{-1}\pi_k(-\alpha)\pi_k(-\alpha^{-1})
  \nonumber\\
  &=&e^{-v(\alpha+\alpha^{-1})}\Big((1-n)\pi_n(-\alpha)\pi_n(-\alpha^{-1})
  \nonumber\\
  &&\quad-\alpha\pi'_n(-\alpha)\pi_n(-\alpha^{-1})
  -\alpha^{-1}\pi_n(-\alpha)\pi'_n(-\alpha^{-1})\Big).
\end{eqnarray}
Defining the dual polynomials $\pi^*_n(z)=z^n\pi_n(z^{-1})$,
the second equality in (\ref{eq:ydef}) is an easy consequence of the
Christoffel-Darboux formula \cite{szegoe67:_orthog_polyn},
\begin{eqnarray}
  \label{eq:CD}
  N_{n}\sum_{k=0}^{n-1}\frac{{\pi_k(a)}\pi_k(b)}{N_k}
  &=&\frac{{\pi^*_{n}(a)}\pi^*_{n}(b)
    -{\pi_{n}(a)}\pi_{n}(b)}{1-{a}b},
\end{eqnarray}
 valid for $a,b\in\C$, $ab\neq1$ and extended by l'Hospital's rule to $ab=1$,
and the trivial relation
\begin{equation}
  \label{eq:prodid}
  {\pi^*_n(z)z^{-1}{\pi^*_n}'(z^{-1})+z\,\pi_n'(z)\pi_n(z^{-1})}
  {=}n\,\pi_n(z)\pi_n(z^{-1})=n\,\pi^*_n(z){\pi^*_n}(z^{-1}). 
\end{equation}
 
Taking only the leading order of $a$ in (\ref{eq:CD}) one obtains the
well-known relations
\begin{eqnarray}
  \label{eq:orthpolrecurs}
  \pi_{n+1}(z)&=&z\,\pi_n(z)+p_{n+1}\pi^*_n(z),
  \nonumber\\[2mm]
  \pi^*_{n+1}(z)&=&z\,p_{n+1}\pi_n(z)+\pi^*_n(z), 
  \\[2mm]
  N_{n+1}&=&N_{n}\big(1-p_{n+1}^2\big)
  \nonumber
\end{eqnarray}
which are closed given $p_n=\pi_n(0)$ for
$n\geq0$. 
For the particular weight function $e^{v(z+z^{-1})}$ one can derive a
nonlinear recursion relation for the $p_n$'s,
\begin{equation}
  \label{eq:PIIp}
  p_n=-\frac{v}n(p_{n+1}+p_{n-1})(1-p_n^2),
\end{equation}
with initial values $p_0=1$,
$p_1=-\frac{I_1(2v)}{I_0(2v)}$.
$I_k(2v)=(2\pi)^{-1}\int_0^{2\pi}e^{ik\theta}e^{2v\cos\theta}d\theta$
is the modified Bessel function of order $k$ and thus $N_0=I_0(2v)$. 
Eq.~(\ref{eq:PIIp}) is the discrete Painlev\'e II equation. It has been
derived in the context of orthogonal polynomials for the first time in
\cite{periwal90:_unitar}, and later on more or less independently in 
\cite{hisakado96:_unitar_matrix_model_painl_iii,tracy99:_random_unitar_matric_permut_painl,baik01:_rieman,borodin01:_discr_painl}.
The differential equations for $\pi_n$, $\pi^*_n$,
\begin{eqnarray}
  \label{eq:diffeq}
  {\pi_n}'(z)&=&
  (n/z+v/z^2-p_{n+1}p_nv/z)\pi_n(z)+(p_{n+1}v/z-p_nv/z^2)\pi^*_n(z)
  \nonumber\\
  {\pi^*_n}'(z)&=&
  (-p_{n+1}v/z+p_nv)\pi_n(z)+(-v+p_{n+1}p_nv/z)\pi^*_n(z),
\end{eqnarray}
can be shown to hold by a tedious but straightforward induction, using
(\ref{eq:orthpolrecurs}) 
and (\ref{eq:PIIp}). They are implicitly derived in
\cite{baik01:_rieman}, from which we learned their actual form. In
\cite{ismail01} an integral expression is obtained for the derivative
of orthogonal polynomials on the circle  with respect to (up to
some technical conditions) an arbitrary weight 
function. Specializing to the weight $e^{v(z+z^{-1})}$ results in a
differential-difference equation equivalent to (\ref{eq:diffeq}).

Of course, the mean of the probability distribution
$F_{x,t}(n)-F_{x,t}(n-1)$ is $2t$ and its variance, the correlation function
(\ref{eq:pngcorr}), is given by
\begin{equation}
  \label{eq:pngcorr2}
  C(x,t)=\sum_{n\geq0}\big(2(n-2t)-1\big)F_{x,t}(n).
\end{equation}
Thus to establish (\ref{eq:generalscalingform}), one has to understand
the scaling properties of the distribution function $F_{x,t}(n)$. Let
us  introduce the new variables $s$, $y$ defined by 
\begin{eqnarray}
  \label{eq:scaledvars1}
  n&=&2v+v^{1/3}s,\\
  \label{eq:scaledvars2}
  x&=&v^{2/3}y,
\end{eqnarray}
where $v=\sqrt{t^2-x^2}$ is regarded as fixed when varying $n$ and $x$.
In \cite{baik00:_limit} the different scaling variable $w=\frac12y$ is
used, which leads to a string of factors of $2$, avoided by our
convention. Setting 
\begin{equation}
  \label{eq:Rfromp}
  R_n=-(-1)^np_n,
\end{equation}
we rewrite (\ref{eq:PIIp}) as
\begin{equation}
  \label{eq:dPII}
  R_{n+1}-2R_n+R_{n-1}=\frac{(\frac{n}v-2)R_n+2R_n^3}{1-R_n^2}.
\end{equation}
Under the scaling (\ref{eq:scaledvars2}),
$R_n=v^{-1/3}u\big(v^{-1/3}(n-2v)\big)+\ord(v^{-1})$, it becomes
the Painlev\'e II equation
\begin{equation}
  \label{eq:PII}
  u''(s)=2u(s)^3+s\,u(s),
\end{equation}
in the limit $v\to\infty$. The
starting value $R_0=-1$ is consistent with the left asymptotics
of $u(s)$ only if
\begin{equation}
  \label{eq:uleftas}
  u(s)\sim-\sqrt{-s/2}\quad\mbox{as $s\to-\infty$},
\end{equation}
which singles out the Hastings-McLeod solution to (\ref{eq:PII})
\cite{hastings80:_painl_kortew_vries}. This particular solution will
be denoted by $u(s)$ and we conclude that 
\begin{equation}
  \label{eq:ulimR}
  u(s)=\lim_{v\to\infty}v^{1/3}R_{[2v+v^{1/3}s]},
\end{equation}
provided the limit exists (A complete proof is the main
content of \cite{baik99}).
$u(s)<0$ and $u$ has the right asymptotics
\begin{equation}
  \label{eq:rightasympt}
  u(s)\sim-\Ai(s)\quad\mbox{as $s\to\infty$}.
\end{equation}
At this point we can derive the scaling limit for $F(n)$.
It is the GUE Tracy-Widom distribution function
\cite{tracy94:_level_airy_kernel} 
\begin{equation}
  \label{eq:FGUE}
  F_{\text{GUE}}(s)=e^{-V(s)},\quad V(s)=-\int_s^\infty v(x)dx,\,\,\,
  v(s)=\big(u(s)^2+s\big)u(s)^2-u'(s)^2, 
\end{equation}
which appears already as the limiting height
distribution for nonstationary curved selfsimilar growth
\cite{praehofer00:_statis}. Since $v'(s)=u(s)^2$, one has $v^{1/3}\log
N_{[2v+v^{1/3}s]}\to v(s)$ 
and $F([2v+v^{1/3}s])\to F_{\text{GUE}}(s)$ as $v\to\infty$. 

Next we turn to the scaling limit for the orthogonal polynomials. For
it to be nontrivial we set
\begin{eqnarray}
  \label{eq:PQ}
  P_n(\alpha)&=&e^{-v\alpha}\pi^*_n(-\alpha),
  \nonumber\\
  Q_n(\alpha)&=&-e^{-v\alpha}(-1)^n\pi_n(-\alpha).
\end{eqnarray}
(\ref{eq:scaledvars2}) implies $\alpha=1-v^{-1/3}y+\ord(v^{-2/3})$.
Setting $n$ as in (\ref{eq:scaledvars1}), we claim that
\begin{equation}
\label{eq:PQa}
a(s,y)=\lim_{v\to\infty}P_n(\alpha),\quad
b(s,y)=\lim_{v\to\infty}Q_n(\alpha).
\end{equation}
If so, the limit functions $a$, $b$ satisfy the differential equations
\begin{eqnarray}
  \label{eq:abs}
  \partial_sa&=&u\,b,
  \nonumber\\
  \partial_sb&=&u\,a-y\,b,
\end{eqnarray}
as a consequence of (\ref{eq:orthpolrecurs}), and 
\begin{eqnarray}
  \label{eq:aby}
  \partial_ya&=&u^2a-(u'+y\,u)b,
  \nonumber\\
  \partial_yb&=&(u'-y\,u)a+(y^2-s-u^2)b,
\end{eqnarray}
as a consequence of (\ref{eq:diffeq}). From (\ref{eq:orthpolrecurs})
one immediately obtains
$\pi^*_n(-1)=(-1)^n\pi_n(-1)=\prod_{k=1}^n(1-R_k)$. 
One has the limit $\prod_{k=1}^\infty(1-R_k)=e^v$ since
$e^{-v^2+v}N_0\prod_{k=1}^nN_k(1-R_k)^{-1}$ has an interpretation as a
probability distribution function \cite{baik99:_algeb}.
Therefore the initial conditions to (\ref{eq:aby}) are
\begin{equation}
  \label{eq:abinit}
  a(s,0)=-b(s,0)=e^{-U(s)},\quad U(s)=-\int_s^\infty u(x)dx.
\end{equation}
The scaling limit of $G_n(\alpha)$ as defined in Eq.~(\ref{eq:ydef}) is
the function
\begin{eqnarray}
  \label{eq:g}
  G(s,y)&=&\int_{-\infty}^sa(s',y)a(s',-y)ds'
  \nonumber\\
  &=&a(s,-y)\partial_ya(s,y)-b(s,-y)\partial_yb(s,y),
\end{eqnarray}
where the second equality can be verified by differentiation with
respect to $s$ and using the identity 
\begin{equation}
  \label{eq:aoutofb}
  a(s,y)=-b(s,-y)e^{\frac13y^3-sy}, 
\end{equation}
itself being a direct consequence of (\ref{eq:aby}) and
(\ref{eq:abinit}). 
Putting these pieces together we obtain as scaling limit for the
distribution functions $F_{x,t}(n)$,
\begin{equation}
  \label{eq:Fydef}
  F_y(s)=\frac{d}{ds}\big(G(s+y^2,y)F_{\text{GUE}}(s+y^2)\big).
\end{equation}
The shift in (\ref{eq:Fydef}) by $y^2$ comes from the fact that
$F_y(s)$ is evaluated for constant $t=v+\frac12v^{1/3}y+\ord(v^{-1/3})$.

In conclusion we arrive at the scaling function
$g(y)$ as defined in the Introduction. From
(\ref{eq:dimensionlessscalingform}), with $\lambda=\frac12$ and $A=2$
for the PNG model, and (\ref{eq:pngcorr2}) we obtain 
\begin{equation}
  \label{eq:gscaling}
  g(y)=\int s^2dF_y(s).
\end{equation}

As already mentioned, except for (\ref{eq:diffeq}), all our relations
are derived in \cite{baik00:_limit,baik01:_rieman}, and the existence
of limits is proven with Riemann-Hilbert techniques. For completeness
let us collect some more properties of $a$ shown in
\cite{baik00:_limit}: 
\begin{eqnarray}
  \label{eq:abprops}
  a(s,y)&\to&1,\quad\mbox{as $s\to+\infty$},
  \nonumber\\
  a(s,y)&\to&0,\quad\mbox{as $s\to-\infty$},
  \nonumber\\
  a\big((2y)^{1/2}x+y^2,y\big)&\to&1,\quad\mbox{as $y\to+\infty$},
  \nonumber\\
  a\big((-2y)^{1/2}x+y^2,y\big)&\to&\frac1{(2\pi)^{1/2}}\int_{-\infty}^x
  e^{-\frac12\xi^2}d\xi,\quad\mbox{as $y\to-\infty$}.
\end{eqnarray}
Therefore $F_y(s)$ is asymptotically Gaussian and we recover
$g(y)\simeq2|y|$ for large $y$.

\section{Numerical determination of the scaling function}
\setcounter{equation}{0}
\label{sec:num}

The key object in determining the scaling functions $g(y)$,
$f(y)$ is 
the Hastings-McLeod solution \cite{hastings80:_painl_kortew_vries} to
Painlev\'e II, $u(s)$, which is the unique solution to
\begin{equation}
  \label{eq:PIIagain}
  u''=2u^3+su
\end{equation}
with asymptotic boundary conditions (\ref{eq:uleftas}) and
(\ref{eq:rightasympt}). Tracy and Widom
\cite{tracy94:_level_airy_kernel,tracy99} integrate (\ref{eq:PIIagain}) 
numerically with conventional differential equation solvers using the 
known asymptotics at $s=\pm\infty$. 
The precision achieved with this technique does not suffice for our
purposes, since we
need $u(s)$ as starting values (\ref{eq:abinit}) for the differential
equations (\ref{eq:aby}). We develop here a different method to obtain
$u(s)$, in principle with arbitrary precision.
Next the functions $a(s,y)$ and $b(s,y)$ have to be determined, which
directly leads to values for the distribution functions $F_y(s)$. They
have to be further integrated with respect to $s$ in order to obtain
their variance, which is the desired scaling function $g(y)$. The
Taylor expansion method to be explained intrinsically produces not only
function values at a point but also higher derivatives. Therefore we
obtain $f(y)$ not by numerically differentiating $g(y)$ but
rather by direct calculation via the knowledge of
$\partial^2_yF_y(s)$. 

In a first step,
to obtain reliable approximations to the Hastings-McLeod solution, we
need to guess its initial data at some finite $s_0$ by using asymptotic 
expansions around $\pm\infty$. Any initial data $u(s_0)=u_0$,
$u'(s_0)=u_1$ give rise to a maximal solution, $\tilde u(s)$ of
(\ref{eq:PIIagain}), 
which admits analytic continuation to a meromorphic function on
$\C$. The only essential singularity for Painlev\'e II solutions lies
at $\infty$. If $\tilde u(s)$ is close to $u(s)$ we can estimate the
difference $\Delta(s)=\tilde u(s)-u(s)$ by linearizing
(\ref{eq:PIIagain}) around $u(s)$. One obtains that
$\Delta(s)/\Delta(0)$ is of order $\exp(D(s))$ if
$(\Delta(s),\Delta'(s))$ is in the 
unstable subspace and of order $\exp(-D(s))$ for the stable
subspace, where  $D(s)\approx {-\frac13(-2s)^{3/2}}$ for $s\ll0$ and
$\approx{\frac23s^{3/2}}$ for $s\gg0$. Note that on the exponential
scale we are looking at, derivation with respect to $s$ leaves
invariant the order. Therefore the exponential orders of $\Delta(s)$
and $\Delta'(s)$ are the same. Since generically initial values at $s_0$ have a
component in both subspaces one obtains that $\Delta(s)/\Delta(s_0)$ is
of order $\exp(|D(s)-D(s_0)|)$ in the range of validity of the
linear approximation, $|\Delta(s)|\ll |u(s)|$.

It turns out that the left asymptotic expansion
in $(-s)^{-1/2}$, optimally truncated at large negative $s_0$, gives
rise to initial values with $\Delta(s_0)$ only of order
$\exp({-\frac13(-2s_0)^{3/2}})$. Thus control of the approximation
always breaks 
down near $s=0$. Approximations of the right
asymptotics on the other hand allow a, in principle, arbitrary
precision on any given finite interval. 

For $s\to\infty$ the deviations of $u(s)$ from the Airy function can
be expanded in an alternating asymptotic power series with
exponentially small prefactor,
\begin{equation}
  \label{eq:rightapproximant}
  u_{\text{right},n}(s)=-\Ai(s)-\frac{e^{-2s^{3/2}}}{32\pi^{3/2}s^{7/4}}
  \sum_{k=0}^n\frac{(-1)^ka_k}{(\frac23s^{3/2})^k}.
\end{equation}
The coefficients are  $a_0=1$, $a_1=\frac{23}{24}$,
$a_2=\frac{1493}{1152}$, $\dots$, and can be obtained via the
recursion relation
\begin{equation}
  \label{eq:rightcoeffs2}
  a_n=\Ai^{(3)}_n+\textstyle\frac34n\,a_{n-1}
  -\frac18(n-\frac16)(n-\frac56)a_{n-2}\quad\mbox{for }n\geq0 
\end{equation}
with initial conditions $a_{-1}=a_{-2}=0$.  
\begin{equation}
  \label{eq:airy3}
  \Ai^{(3)}_n=\sum_{0\leq k\leq l\leq n}\Ai_{n-l}\Ai_{l-k}\Ai_{k}.
\end{equation}
are the coefficients in the asymptotic expansion of $\Ai(x)^3$ and
\begin{equation}
  \label{eq:Airecursion}
  \Ai_n=\frac{(6n-1)(6n-5)}{72n}\Ai_{n-1},\qquad \Ai_0=1
\end{equation}
are the coefficients of the asymptotic expansion of the Airy function
itself \cite{abramowitz84:_pocket_mathem_funct},
\begin{equation}
  \label{eq:Aiexpansion}
  \Ai(s)\sim\frac{e^{-\frac23s^{3/2}}}{2\sqrt{\pi}s^{1/4}}
  \sum_{n\geq0}\frac{(-1)^n}{(\frac23s^{3/2})^n}\Ai_n.
\end{equation}

Empirically we observe that for $s_0\gg0$  the optimal truncation in
(\ref{eq:rightapproximant}) is $n\approx \frac43s_0^{3/2}$ leading to
an exponentially improved (relative) precision
\begin{equation}
  \label{eq:relativeerror}
  \left|\frac{u_{\text{right},n}(s_0)-u(s_0)}{u(s_0)}\right|\approx
  \exp({-\textstyle\frac83s_0^{3/2}}).
\end{equation}
The linear perturbation argument for $\tilde u(s)$ with initial
values  $\tilde u(s_0)=u_{\text{right},n}(s_0)$,  $\tilde
u'(s_0)=u_{\text{right},n}'(s_0)$, $n=[\frac43s_0^{3/2}]$, is now valid
for a slightly smaller interval than $[-2s_0,3^{3/2}s_0]$.
For example, by choosing the interval $[-2^{1/3}s_0,s_0]$ and gluing
$\tilde u(s)$ at the boundaries smoothly to the optimally truncated asymptotic
expansions, the maximal relative error of $\tilde u(s)$ with respect to the
Hastings-McLeod solution is of order $e^{-4/3s_0^{3/2}}$.
For our purpose it turns out that we do not need values with
$s<-20$. On the other hand, to access large values of $y$ in
(\ref{eq:Fydef}), because of the shift in (\ref{eq:Fydef}), we need 
$u(s)$ for large $s$ with high precision. $u_{\text{right},n}(s)$ is
numerically 
costly to evaluate for large $s$, so we finally choose $s_0=100$ and
integrate in the interval $[-20,200]$. This requires a maximal working
precision of $~1500$ digits and, given the integration of
(\ref{eq:PIIagain}) is precise enough, $\tilde u(s)$ and $u(s)$
coincide in the first $~1000$ digits for $s\in[-20,115]$ and still up to 
$50$ digits at $s=200$, where $u(s)\approx 10^{-820}$. In the sequel
we drop the distinction between $u(s)$ and its numerical approximation.
The arithmetic computing is done partially with
Mathematica$^{\mbox{\tiny\textregistered}}$ and for the
computationally intensive tasks with the C++-based multiprecision
package {\tt MPFUN++} \cite{chatterjee00:_mpfun_c_multip_system}.

To solve initial value problems for ordinary differential equations
highly sophisticated iteration schemes are available, like
Runge-Kutta, Adams-Bash\/ford and multi-step methods. For arbitrary
high (but fixed) precision results, all these methods become
ineffective, since the step size is a decreasing function of the
required precision goal for the solution and tends to become
ineffectively small. The only remaining choice is to Taylor expand the
solution at a given point. The step size is limited by the radius of
convergence only and the precision is controlled by the error made in
truncating the Taylor series at some order
\cite{barton71:_taylor_series_method_ordin_differ_equat}.

$u(s)$ is expanded at $s_0$ as
\begin{equation}
  \label{eq:PIIexpansion}
  u(s)=\sum_{n\geq0} u_n(s-s_0)^n.
\end{equation}
For the Painlev\'e II equation the expansion coefficients  $u_n$
at $s_0$ are determined by $u_0=u(s_0)$, $u_1=u'(s_0)$ and 
\begin{equation}
  \label{eq:PIItaylorrecursion}
  u_{n+2}=\frac{2u^{(3)}_n+s_0\,u_n+u_{n-1}}{(n+2)(n+1)},
\end{equation}
where $u^{(k)}_n=\sum_{j=0}^nu_{n-j}u^{(k-1)}_j$ are the expansion
coefficients of $u(s)^k$ at $s_0$, $u^{(1)}_n=u_n$. We include
the factorial into the expansion
coefficients instead of taking the bare Taylor coefficients, in order to
reduce the workload from multiplications by binomials
when multiplying two expansions numerically. 

Numerically we find that the Hastings-McLeod solution does not have
any pole in a strip $|\text{Im}(s)|<2.9$. To have a safety margin we
choose a step size one for the extrapolation of the expansion
(\ref{eq:PIIexpansion}).

We take the starting values  $u(s_0)$, $u'(s_0)$ from
(\ref{eq:rightapproximant}) at $s_0=100$. The coefficients of the
functions $U(s)$, $V(s)$, see (\ref{eq:FGUE}) and (\ref{eq:abinit}),
when expanded around $s_0$, as in (\ref{eq:PIIexpansion}), are given by  
\begin{equation}
  \label{eq:UandVexpansion}
  U_{n+1}=\frac{u_n}{n+1},\quad
  V_{n+2}=\frac{u^{(2)}_n}{(n+2)(n+1)},\quad n\geq0,
\end{equation}
and $V_1=u_0^4-u_1^2+s_0u_0^2$, leaving unspecified the yet unknown
integration constants $U_0=U(s_0)$ and $V_0=V(s_0)$. The higher expansion
coefficients of $u$, $u'$, $U$, and $V$ are independent of the values
for $U_0$ and $V_0$ and are calculated with the recursion
relations (\ref{eq:PIItaylorrecursion}) and
(\ref{eq:UandVexpansion}). The values of 
$u$, $u'$, $U$, and $V$ at $s=s_0\pm1$ are extrapolated, the
expansion coefficients at $s_0\pm1$ iterated. Then values are calculated at  
$s=s_0\pm2$ by extrapolation, and so on. A posteriori we
assign to $U(s_0)$ and $V(s_0)$ values, such that $U(200)=0=V(200)$.
The numerical errors from iterating
(\ref{eq:PIItaylorrecursion}) and from truncating
(\ref{eq:PIIexpansion}) can be neglected compared to the uncertainty
originating from the 
initial conditions. At the end of this first step we have at our
disposal the expansion coefficients for $u,u',U,V$ at the integers in
the interval $[-20,200]$. 
For the convenience of the interested reader let us just state the
values at $s=0$ up to $50$ digits,
\begin{eqnarray}
  \label{eq:atzero}
  u(0)&=&\mbox{\small\tt-0.367061551548078427747792113175610961512192053613139},
  \nonumber\\
  u'(0)&=&\mbox{\small\tt0.295372105447550054557007047310237988227233798735629},
  \nonumber\\
  U(0)&=&\mbox{\small\tt0.336960697930551393597884426960964843885993886628226},
  \nonumber\\
  V(0)&=&
  \mbox{\small\tt0.0311059853063123536659591008775670005642241689547838},
  \nonumber
\end{eqnarray}
which might be used as starting values for a quick conventional   
integration of Painlev\'e II to reproduce parts of our results with
much less effort but also less precision. Tables can be found at
\cite{praehofer:_y}. 

The next step is to determine $a(s,y)$, $b(s,y)$ at
$s_0\in\{-20,\dots,200\}$ in the interval $y\in[-9,9]$ employing
(\ref{eq:aby}) and (\ref{eq:abinit}). Setting
\begin{eqnarray}
  \label{eq:abexpansion}
  a(s,y)=\sum_{m,n\geq0}a_{m,n}(s-s_0)^m(y-y_0)^n,
  \nonumber\\
  b(s,y)=\sum_{m,n\geq0}b_{m,n}(s-s_0)^m(y-y_0)^n,
\end{eqnarray}
(\ref{eq:aby}) becomes a recursion relation for the expansion
coefficients,
\begin{eqnarray}
  \label{eq:abyrecursion}
  a_{m,n+1}&=&\frac1{n+1}\sum_{k=0}^m \big(u^{(2)}_ka_{m-k,n}
  -(k+1)u_{k+1}b_{m-k,n}-u_{k}b_{m-k,n-1}\big),
  \nonumber\\
  b_{m,n+1}&=&\frac1{n+1}\Big(b_{m,n-2}-b_{m-1,n}
  \nonumber\\
  &&+\sum_{k=0}^m \big(-u^{(2)}_kb_{m-k,n}
  +(k+1)u_{k+1}a_{m-k,n}-u_{k}a_{m-k,n-1}\big)\Big), 
  \nonumber\\
\end{eqnarray}
$n\geq0$, allowing one to determine $a_{0,n}$, $b_{0,n}$, $n\geq0$
upon the knowledge of $a_{0,0}$, $b_{0,0}$.
We integrate along $\pm y$ with an extrapolation step size of
$\frac18$. 
From (\ref{eq:abs}) one obtains 
the recursions 
\begin{eqnarray}
  \label{eq:absrecursion}
  a_{m+1,n}&=&\frac1{m+1}\sum_{k=0}^m u_k b_{m-k,n}
  \nonumber\\
  b_{m+1,n}&=&\frac1{m+1}\Big(-b_{m,n-1}+\sum_{k=0}^m u_k a_{m-k,n}\Big).
\end{eqnarray}
The expansion coefficients $G_{m,n}$ of $G(s,y)$ at $(s_0,y_0)$,
are determined from (\ref{eq:g}) as
\begin{equation}
  \label{eq:gexpansion}
  G_{m,n}=(n+1)\big(a^{-}_{m,n}a_{m,n+1}-b^{-}_{m,n}b_{m,n+1}\big)
\end{equation}
where $a^-_{m,n}$, $b^-_{m,n}$ are the corresponding expansion
coefficients of $a$ and $b$ at $(s_0,-y_0)$.

To finally determine $g(y)$ and its derivatives we write 
\begin{eqnarray}
  \label{eq:gsplit}
  g^{(n)}(y_0)&=&\frac{d^n}{dy_0^n}\sum_{s_0\in\Z}\int_{s_0}^{s_0+1}(s-y_0^2)^2
  \frac{d^2}{ds^2}\big(G(s,y_0)F_{\text{GUE}}(s)\big)ds
  \nonumber\\
  &=&\sum_{s_0\in\Z}n!\sum_{m\geq1}c_{m,n}.
\end{eqnarray}
$c_{m,n}$ are the expansion coefficients of
$(s,y)\mapsto\int_{s_0}^{s}(r-y^2)^2\frac{d^2}{dr^2}\big(G(r,y)
F_{\text{GUE}}(r)\big)dr$
at $(s_0,y_0)$,
\begin{equation}
  \label{eq:cmn}
  c_{m,n}=(m-2)(GF)_{m-1,n}-2m(GF)_{m,n-2}
  +\textstyle\frac{m(m+1)}{m-1}(GF)_{m+1,n-4}.
\end{equation}
Here 
\begin{equation}
  \label{eq:gfmn}
  (GF)_{m,n}=\sum_{k=0}^m F_kG_{m-k,n}
\end{equation}
are the expansion coefficients of $G(s,y)F_{\text{GUE}}(s)$
and $F_n=-\sum_{k=1}^n\frac{k}{n}V_kF_{n-k}$ are the expansion
coefficients of $F_{\text{GUE}}$.
Numerically the sum over $s_0$ in (\ref{eq:gsplit}) is truncated to
values inside $[-15,200]$, since outside contributions turn out to be
negligible at the chosen precision goal. After accomplishing this
program we keep values for $g(y)$ at
$y\in\frac1{128}\Z\cap[-9,9]$ and for $g^{(n)}(y)$, $n=0,\dots,4$, at
$y\in\frac1{8}\Z\cap[-9,9]$ with an accuracy of about $100$ digits (a
table in ASCII format is available online at \cite{praehofer:_y}).
For interpolating these values we used the {\tt
  Interpolation}-function of the
Mathematica$^{\mbox{\tiny\textregistered}}$ package yielding best results
due to the high precision data with an interpolation order of $57$.

\section{Discussion of the scaling function}
\label{sec:discussion}

There have been numerous attempts to approximately determine $g(\cdot)$
\cite{beijeren85:_exces,hwa91:_exact,tang92:_stead,frey96:_mode_burger,fogedby01:_scalin_burger}.
For historical reasons a different scaling function, $F(\cdot)$, is
analyzed in some of these works. The relation to our scaling function
$g(\cdot)$ is 
\begin{eqnarray}
  \label{eq:Ftog}
  F(\xi)=(\xi/2)^{2/3}\,g\left((2\xi^2)^{-1/3}\right),\quad\mbox{resp.}
  \quad
  g(y)=2y\,F\big(1/({2^{1/2}y^{3/2}})\big).
\end{eqnarray}
Note that by (\ref{eq:dimensionlessscalingform}) the large $y$
behavior of $g$ is fixed by definition as $g(y)\sim2|y|$. The special
value $g(0)=1.1503944782594709729961$ is the Baik-Rains constant
\cite{baik00:_limit,praehofer00:_univer}. In the literature the
universal amplitude ratio $R_G=2^{-2/3}g(0)=0.7247031092$ and the
universal coupling constant $g^*=g(0)^{-3/2}=0.810456700$ have been
investigated. Approximate values have been determined by means of 
Monte-Carlo simulations for the single step model
\cite{tang92:_stead}, numerically within a mode-coupling
approximation
\cite{hwa91:_exact,frey96:_mode_burger,
moore01:_numer_solut_mode_coupl_equat},
and even experimentally for slowly combusting paper
\cite{myllys01:_kinet} yielding estimates for $g(0)$ within reasonable
ranges around the (numerically) exact value indicated.

\begin{figure}[ht]
  \centering
  \includegraphics[width=10cm]{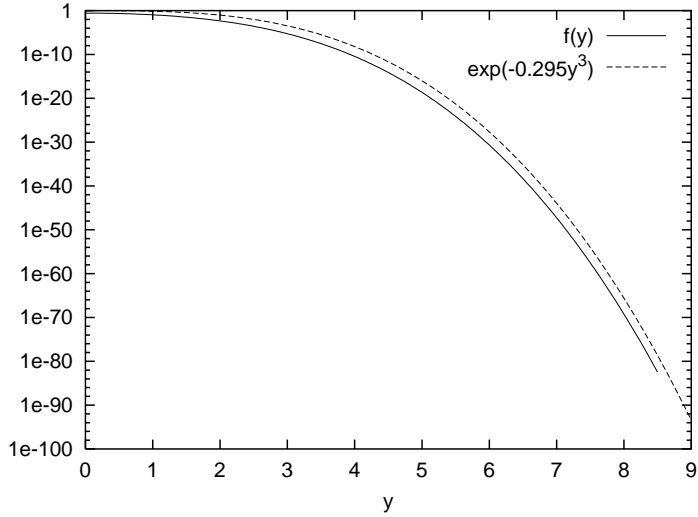}
  \caption{\it The scaling function $f(y)$ versus $y$ in a semilogarithmic
  plot. The dotted line $\exp(-0.295|y|^3)$ is drawn as a guide to the
  eye for the large $y$ asymptotics of $f$.} 
  \label{fig:fy}
\end{figure}

In Figure \ref{fig:fy} the scaling function $f(y)=\frac14g''(y)$ is
shown as determined by the multiprecision expansion method explained
in the previous section. We estimate its large $y$ asymptotics as
\begin{equation}
  \label{eq:fasympt}
  \log f(y)\approx-c|y|^3+o(|y|)\mbox{ for $y\to\infty$}.
\end{equation}
The cubic behavior is very robust and numerical fits yield about
$2.996$--$2.998$ 
quite independently of the assumed nature of the finite size
corrections. The prefactor $c=0.295(5)$ has a relatively high
uncertainty because of the unknown subleading corrections. Even though
unaccessible in nature we estimate the error term, as indicated in
(\ref{eq:fasympt}), to be sublinear or even
only logarithmic from the numerical data. Possibly, the exact
asymptotic behavior could be extracted from a refined asymptotic
analysis of the Riemann-Hilbert problem.

\begin{figure}[ht]
  \centering
  \includegraphics[width=10cm]{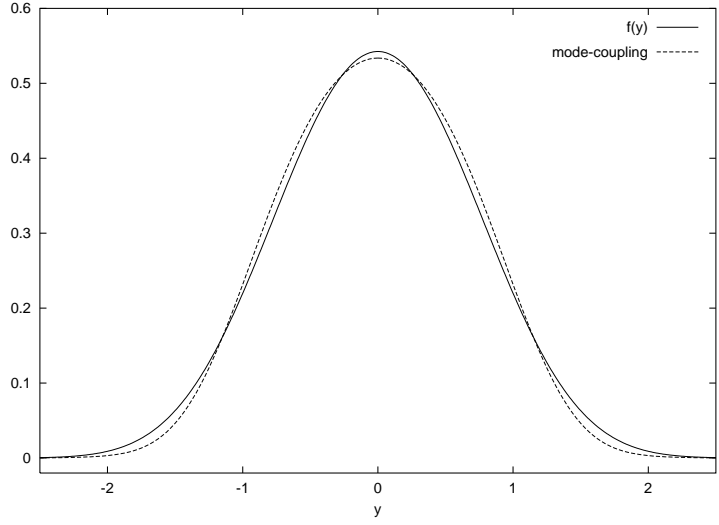}
  \caption{\it The exact scaling function $f(y)$ compared to the mode
  coupling result of Colaiori and Moore
  \cite{moore01:_numer_solut_mode_coupl_equat}(dotted line). Both
  functions are even.} 
  \label{fig:fvsmc}
\end{figure}

Colaiori and Moore \cite{moore01:_numer_solut_mode_coupl_equat,
moore01:_stret_kardar_paris_zhang} tackled the same scaling
function by completely different means. Starting from the continuum
version of the KPZ equation they numerically solved the corresponding
mode-coupling equation \cite{beijeren85:_exces,frey96:_mode_burger},
which contains an uncontrolled approximation, since diagrams which
would renormalize the three-point vertex coupling are
neglected. Nevertheless a 
qualitative comparison of their result with the exact scaling function
$f(y)$ shows reasonable similarity, cf.~Figure \ref{fig:fvsmc}. 
Both functions are normalized to integral $1$ by definition. The mode
coupling solution oscillates around $0$ for $|y|>3$, whereas $f(y)>0$
for the exact solution. We do not know whether this is a numerical
artifact or an inherent property of the mode-coupling
equation. On the other hand, the second moments are reasonably
close 
together, $0.510523$ for $f(y)$, and  $0.4638$ for the mode-coupling 
approximation. So is the value of the Baik-Rains constant
$g(0)=2\int|y|f(y)dy$ for which mode-coupling predicts the value
$1.1137$. 

\begin{figure}[ht]
  \centering
  \includegraphics[width=10cm]{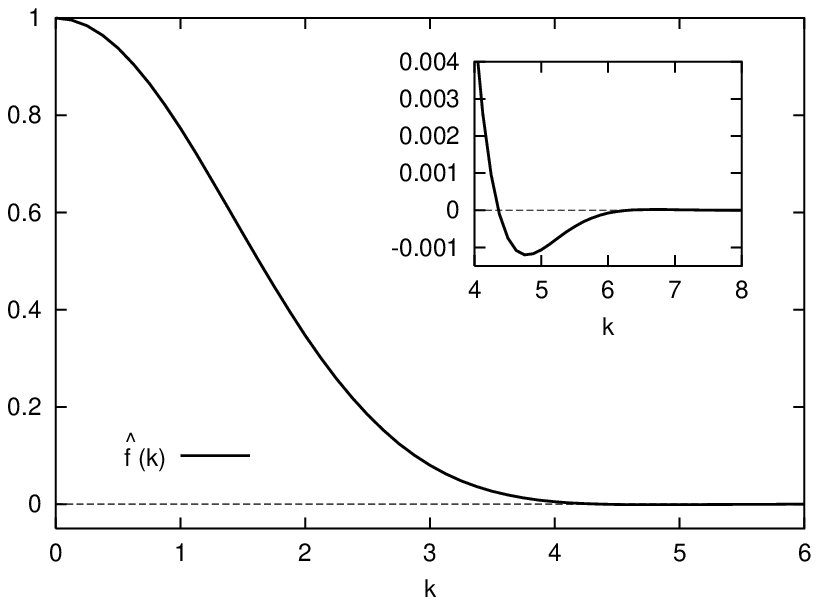}
  \caption{\it The Fourier transform $\widehat{f}(k)$ of the scaling
    function $f(y)$.}
  \label{fig:fk}
\end{figure}

From the solution to the mode-coupling equations one does not directly
obtain $f(y)$, but rather its Fourier transform. The function
$G(\tau)$ from \cite{moore01:_numer_solut_mode_coupl_equat} is defined
through  
\begin{equation}
  \label{eq:Gfromf}
  G(k^{3/2}/2^{7/2})=\widehat{f}(k)=2\int_0^\infty\cos(ky)f(y)dy.
\end{equation}
Moore and Colaiori predict a stretched exponential decay of $G(\tau)$
as  $\propto\exp(-c|\tau|^{2/3})$
\cite{moore01:_stret_kardar_paris_zhang} and numerically find a
superimposed oscillatory behavior on the scale $|\tau|^{2/3}$
\cite{moore01:_numer_solut_mode_coupl_equat}. In Figure \ref{fig:fk}
\begin{figure}[tb]
  \centering
  \includegraphics[width=10cm]{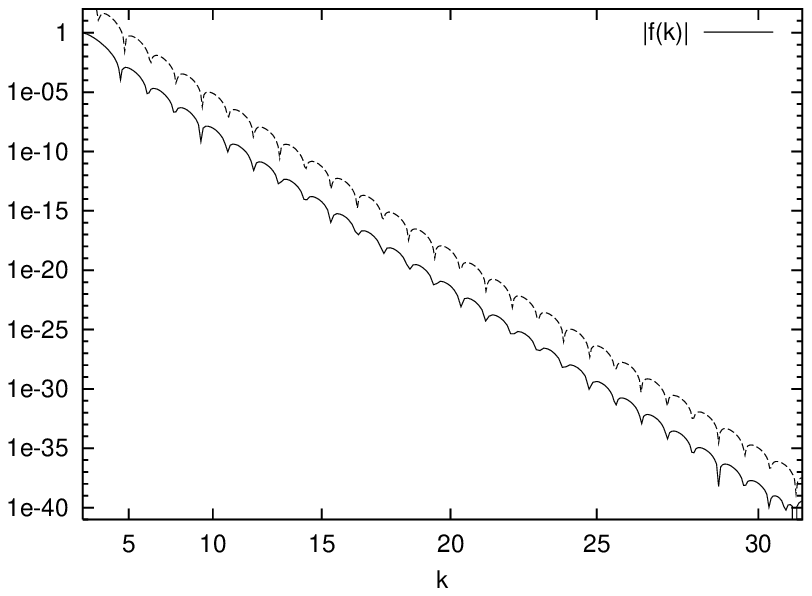}
  \caption{\it The modulus of $\widehat{f}(k)$ on a semilogarithmic
  scale. The dotted line is a heuristic fit, shifted by a factor
  $1000$ for visibility.} 
  \label{fig:fklog}
\end{figure}
$\widehat{f}(k)$ is plotted as obtained by  a numerical Fourier
transform of $f(y)$. Indeed it exhibits an oscillatory behavior as can be
seen in Figure \ref{fig:fklog} where the modulus of $\widehat{f}(k)$
is shown on a semilogarithmic scale. The dotted line in the plot is
the modulus of the function 
\begin{equation}
  \label{eq:fittofhat}
  10.9k^{-9/4}\sin(\textstyle\frac12k^{3/2}-1.937)e^{-\frac12k^{3/2}},
\end{equation}
shifted by a factor of $1000$ for visibility, which fits
$\widehat{f}(k)$ very well in phase and amplitude for 
$k\gtrapprox15$. This behavior is not in accordance with the
results of Colaiori and Moore, since the oscillations and the
exponential decay of $G(\tau)$ for the exact solution are apparently
on the scale $\tau$ and not $\tau^{2/3}$.

\begin{figure}[htb]
  \centering
  \includegraphics[width=12cm]{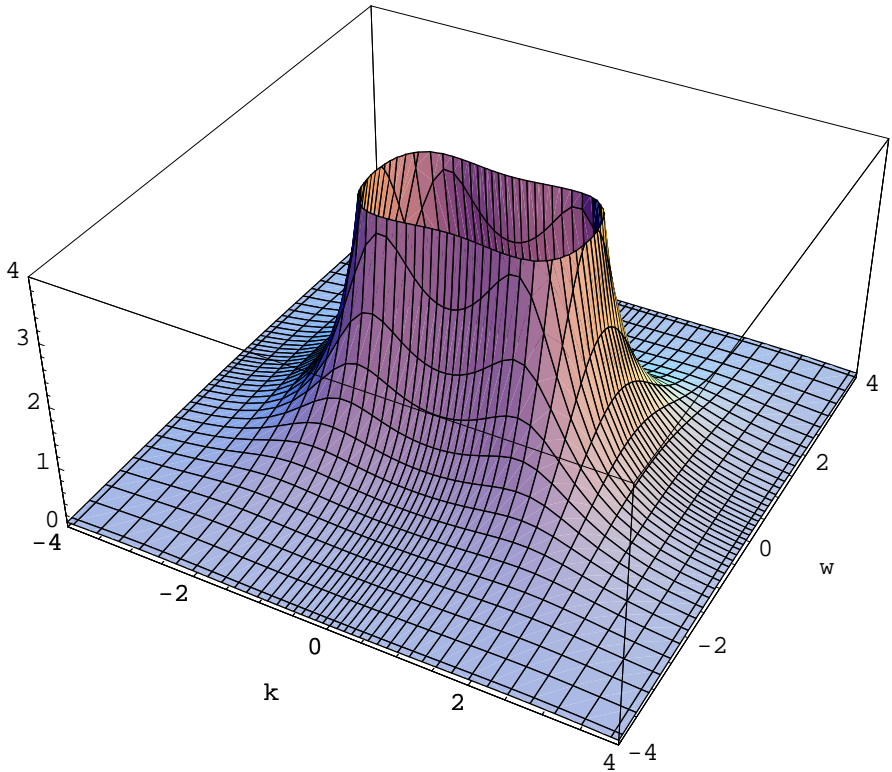}
  \caption{\it The correlation function $C^{\text{KPZ}}(k,\omega)$ in Fourier space.}
  \label{fig:Ckw}
\end{figure}

Note that $\widehat{f}(k)$ is the scaling function for the
intermediate structure function 
\begin{equation}
  \label{eq:structure}
  S(k,t)=\int dx e^{ikx}S(x,t)\simeq2\widehat{f}(t^{2/3}k).
\end{equation}
By Fourier transforming with respect to $t$ we determine the
dynamical structure function, 
\begin{equation}
  \label{eq:fullstructure}
  S(k,\omega)=\int dx\,dt e^{i(kx+\omega
  t)}S(x,t)\simeq 2k^{-3/2}\stackrel{\circ}{f}(\omega/k^{3/2}), 
\end{equation}
where
\begin{eqnarray}
  \label{eq:fhathat}
  \stackrel{\circ}{f}(\tau)=\int ds\, e^{i\tau
  s}\widehat{f}(s^{2/3})=2\int_0^\infty
  dy\,\tau^{-1}L'(y/\tau^{2/3})f(y)
\end{eqnarray}
and $L$ has the convenient representation
\begin{equation}
  \label{eq:Ldef}
  L(\kappa)=2\cdot3^{2/3}\Ai(-3^{-4/3}\kappa^2)\sin(2\kappa^3/27).
\end{equation}
The correlation function (\ref{eq:pngcorr}) in Fourier
space is given by
\begin{equation}
  \label{eq:CfromS}
  C(k,\omega)=2k^{-2}S(k,\omega)
  \sim C^{\text{KPZ}}(k,\omega)
  \stackrel{\text{def}}{=}4k^{-7/2}\stackrel{\circ}{f}(\omega/k^{3/2}),
\end{equation}
describing the asymptotic behavior at $k,\omega=0$. Note that
$C(k,\omega)>0$ by definition, since $\langle
h_{k,\omega}h_{k',\omega'}\rangle=\delta_{k,-k'}
\delta_{\omega,-\omega'}C(k,\omega)$ for $(k,\omega)\neq(0,0)$.
The anomalous scaling behavior in real space is reflected by the
exponents for the divergence of $C^{\text{KPZ}}(k,\omega)$ at
$k=\omega=0$. In the linear case, the Edwards-Wilkinson equation
$\lambda=0$ in (\ref{eq:KPZ1d}), one easily obtains  
\begin{equation}
  \label{eq:CEW}
  C^{\text{EW}}(k,\omega)=\frac{D}{\omega^2+\nu^2k^4}.
\end{equation}
A 3d-plot of $C^{\text{KPZ}}(k,\omega)$ is shown in Figure \ref{fig:Ckw}. 
Its striking features are the smooth behavior
away from $k,\omega=0$, especially on the lines where $k=0$ and
$\omega=0$ and the two symmetric maxima of $k\mapsto
C^{\text{KPZ}}(k,\omega)$ for constant $\omega$. Our numerical data
yield for the singular behavior at $k=0$, $\omega=0$, 
\begin{eqnarray}
  \label{eq:Casympt}
  C^{\text{KPZ}}(k,\omega)&=&\omega^{-7/3}
  \big(2.10565(1)+0.85(1)\,k^2w^{-4/3}+\ord(k^4\omega^{-8/3})\big),
  \nonumber\\
  &=&k^{-7/2}\big(19.4443(1)-52.5281(1)\,\omega^2k^{-3}
  +\ord(\omega^4 k^{-6})\big).
\end{eqnarray}


\section{Conclusions and Outlook}
\label{sec:conclusions}
\setcounter{equation}{0}
For systems close to equilibrium many properties valid in generality
rely on detailed balance, amongst them in particular the link between
correlation and response functions. The KPZ equation does not satisfy
detailed balance, since the growth is directed. However, it has been
speculated that in $1+1$ dimensions detailed balance is recovered in
the scaling regime. With our exact scaling function at hand, such a 
claim can be tested.

Detailed balance implies that the eigenvalues of the 
generator in the master equation lie on the negative real axis. Thus
autocorrelations in the 
form $\langle X(t)X(0)\rangle$ can be written as the Laplace transform
of a positive measure. The structure function $S(k,t)$ at fixed $k$ is
such an autocorrelation. Using the scaling form (\ref{eq:structure})
detailed balance would imply 
\begin{equation}
  \label{eq:detbal}
  S(k,t)=\int_{-\infty}^0\nu\big(|k|^{-3/2}d\lambda\big)e^{\lambda|t|}
\end{equation}
with $\nu(d\lambda)\geq0$. In particular $S(k,t)\geq0$. From
(\ref{eq:fittofhat}) we know that $S(k,t)$ oscillates around zero.
Definitely, at $|k|\approx5$ there is a negative dip,
cf.~Fig.~\ref{fig:fk}. Thus (\ref{eq:detbal}) cannot be correct.

The Bethe ansatz \cite{gwa92:_six_hamil,gwa92:_bethe_burger} indicates
that, for large system size,  the density of states is concentrated on
an arc touching $0$. If 
so, the integration in (\ref{eq:detbal}) would have to be replaced by
a corresponding line integral. It is not clear to us how to extract
from the numerical knowledge of $S(k,t)$ such a representation.

Our main result is the exact scaling function $f$, see Figure
\ref{fig:fvsmc}, for the two-point function of the stationary KPZ equation in $1+1$ dimensions. ``Exact'' must be qualified in two
respects. Firstly $f$ is given indirectly through the solution of
certain differential equations, which can be solved numerically only
with considerable effort. The errors are well controlled,
however. Secondly, we rely on universality, in the sense that the
scaling function is derived for the PNG model, which is one rather
particular model within the KPZ universality class. Of course, it
would be most welcome to establish the scaling limit also for other
models in this class.

The KPZ equation (\ref{eq:KPZ1d}) is a two-dimensional field theory
and, in spirit, belongs to the same family as two-dimensional
models of equilibrium statistical mechanics, one-dimensional quantum
spin chains, and other $(1+1)$-dimensional quantum field theories at
zero temperature. While in the latter cases, there are a number of
models for which the two-point function can be computed, in the
dynamical context such solutions are scarce. In addition, the KPZ
equation does not satisfy the condition of detailed balance. Such
nonreversible models are known to be difficult and we believe that the
PNG model is the first one in the list of exact solutions,
disregarding noninteracting field theories.

For the nonstationary KPZ equation with a macroscopic profile of
nonzero curvature the analogue of $F_0$ is the Tracy-Widom
distribution function. In that case the full statistics of $x\mapsto
h(x,t)$ for large but fixed $t$ is available
\cite{praehofer01:_scale_invar_png_dropl_airy_proces}. It is
conceivable that an extension of the techniques used there also admits
a more detailed study of, say, the joint distribution of
$h(x,t)-h(0,0)$, $h(x',t)-h(0,0)$. On the other hand the joint
distribution of $h(0,t)-h(0,0)$, $h(0,t')-h(0,0)$ does not seem to be
accessible. In the representation through the
directed polymers it means that space-like points, even several of
them, can be handled, whereas time-like points remain a challenge.
\medskip\\
{\bf Acknowledgments.} We greatly enjoyed the collaboration with
Jinho Baik at the early stage of this project and are most grateful
for his important input. We also thank Mike~Moore for providing
us with the numerical solution of the mode-coupling equation.



\end{document}